\documentclass[twocolumn]{aa}
\usepackage{graphicx}
\usepackage{multirow}
\usepackage{enumitem} % for no indent in itemize 
\usepackage[]{algorithm2e}

\usepackage{xcolor}
\usepackage{subfigure}
\usepackage[normalem]{ulem}
\usepackage{mathtools}
\usepackage{amsmath}

\usepackage{txfonts}
\usepackage{textcomp} % for \textmu

\usepackage{ulem}

\begin{document}

   \title{On-sky validation of image-based adaptive optics wavefront sensor referencing}

   %\subtitle{I. Overviewing the $\kappa$-mechanism}

\author{Nour Skaf\inst{\ref{inst:LESIA},\ref{inst:Subaru},\ref{inst:UCL}} \and%
    Olivier Guyon\inst{\ref{inst:Subaru},\ref{inst:ABC},\ref{inst:UoA}, \ref{inst:UoAO}}
	Éric Gendron\inst{\ref{inst:LESIA}} \and%
	Kyohoon Ahn\inst{\ref{inst:Subaru}} \and%
	Arielle Bertrou-Cantou\inst{\ref{inst:LESIA}} \and%
	Anthony Boccaletti\inst{\ref{inst:LESIA}} \and%
	Jesse Cranney\inst{\ref{inst:ANU}} \and%
	Thayne Currie\inst{\ref{inst:Subaru},\ref{inst:Ames}}\and%
	Vincent Deo\inst{\ref{inst:Subaru}} \and%
	Billy Edwards\inst{\ref{inst:UCL},\ref{inst:CEA}} \and%
	Florian Ferreira\inst{\ref{inst:LESIA}} \and%
	Damien Gratadour\inst{\ref{inst:LESIA},\ref{inst:ANU}}\and\\%
	Julien Lozi\inst{\ref{inst:Subaru}}\and%
	Barnaby Norris\inst{\ref{inst:SydneyUni}} \and 
	Arnaud Sevin\inst{\ref{inst:LESIA}} \and%
	Fabrice Vidal\inst{\ref{inst:LESIA}} \and%
	Sébastien Vievard\inst{\ref{inst:Subaru},\ref{inst:ABC}}
	}%
	
\institute{LESIA, Observatoire de Paris, Univ.~PSL, CNRS, Sorbonne Univ., Univ.~de Paris, 5 pl. Jules Janssen, 92195 Meudon, France\label{inst:LESIA}%
\and %
National Astronomical Observatory of Japan, Subaru Telescope, 650 North A'oh\=ok\=u Place, Hilo, HI 96720, U.S.A.\label{inst:Subaru} %
\and 
Department of Physics and Astronomy, University College London, London, United Kingdom\label{inst:UCL} %
\and 
Astrobiology Center of NINS, 2-21-1 Osawa, Mitaka, Tokyo 181-8588, Japan\label{inst:ABC}
\and
Steward Observatory, University of Arizona, Tucson, AZ 85721, USA\label{inst:UoA}
\and
College of Optical Sciences, University of Arizona, Tucson, AZ 85721, USA\label{inst:UoAO}
\and %
Research School of Astronomy and Astrophysics, Australian National University, Canberra, ACT 2611, Australia\label{inst:ANU} %
\and
NASA-Ames Research Center, Moffett Field, California, USA\label{inst:Ames}
\and
AIM, CEA, CNRS, Universit\'e Paris-Saclay, Universit\'e de Paris, F-91191 Gif-sur-Yvette, France\label{inst:CEA}
\and
Sydney Institute for Astronomy, School of Physics, University of Sydney, NSW 2006\label{inst:SydneyUni}
\\
email: \texttt{nour.skaf@obspm.fr}
}

   \date{Received 10/06/2021; Accepted 04/10/2021}

{

}

  \abstract
   {

   Differentiating between a true exoplanet signal and residual speckle noise is a key challenge in high-contrast imaging (HCI).

   Speckles are due to a combination of fast, slow and static wavefront aberrations introduced by atmospheric turbulence and instrument optics.
   While wavefront control techniques developed over the last decade have shown promise in minimizing fast atmospheric residuals, slow and static aberrations such as non-common path aberrations (NCPAs) remain a key limiting factor for exoplanet detection.

   NCPAs are not seen by the wavefront sensor (WFS) of the adaptive optics (AO) loop, hence the difficulty in correcting them.}
   {
   We propose to improve the identification and rejection of slow and static speckles in AO-corrected images.
   The algorithm DrWHO - Direct Reinforcement Wavefront Heuristic Optimisation -, performs frequent compensation of static and quasi-static aberrations (including NCPAs) to boost image contrast and is applicable to general purpose AO systems as well as HCI systems. 
   }
   {By changing the WFS reference at every iteration of the algorithm (few tens of seconds), DrWHO changes the AO system point of convergence to lead it towards a compensation of the static and slow aberrations. References are calculated using an iterative lucky-imaging approach, where each iteration updates the WFS reference, ultimately favoring high-quality focal plane images. 
   
   }
   {We validate this concept through both numerical simulations and on-sky testing on the SCExAO instrument at the 8.2-m Subaru telescope. Simulations show a rapid convergence towards the correction of 82\% of the NCPAs. On-sky tests are performed over a 10-minute run in the visible (750~nm). We introduce a flux concentration (FC) metric to quantify the point spread function (PSF) quality and measure a  15.7\% improvement compared to the pre-DrWHO image.}
   {The DrWHO algorithm is a robust focal-plane wavefront sensing calibration method that has been successfully demonstrated on sky. It does not rely on a model and does not require wavefront sensor calibration or linearity. It is compatible with different wavefront control methods, and can be further optimized for speed and efficiency. 
   The algorithm is ready to be incorporated in scientific observations, enabling better PSF quality and stability during observations.}


   \maketitle

\section{Introduction} \label{sec:intro}

Over the past 30 years, Adaptive Optics (AO) instrumentation has undergone extensive growth in sophistication and scientific capabilities. Compared to the first astronomical AO experiments with the COME-ON system consisting of a 19-actuator deformable mirror (DM) driven at several hundred Hz \citep{Kern1989,Rousset1990}, current leading extreme AO (ExAO) systems consists of $\approx$2000-actuator DMs driven at over 1 kHz speeds \citep[e.g.][]{GPIMacintosh2014,Beuzit2019}. The high-contrast imaging enabled by these new AO capabilities has led to key scientific breakthroughs, including the first direct image of a planet-mass companion \citep{chauvin2004} and the first imaged system of jovian exoplanets \citep{Marois2008}.  

Some of the current generation of ExAO systems, e.g. the Subaru Coronagraphic Extreme Adaptive Optics (SCExAO) at the Subaru Telescope \citep{Jovanovic2015} and MagAO-X on the Magellan Clay telescope \citep{Males2018MagAOx} are pushing further achievable performance -- e.g. Strehl ratio (SR), contrast and sensitivity \citep[e.g.][]{Vigan2015,Currie2020}--  and serve as technology prototypes for ExAO instruments on 25-40 m extremely large telescopes \citep{Kasper_PCS,TMTwhitepaper_Fitzgerald, GMT_2020}.

\begin{figure}[t]
\centering
\includegraphics[width=0.45\textwidth]{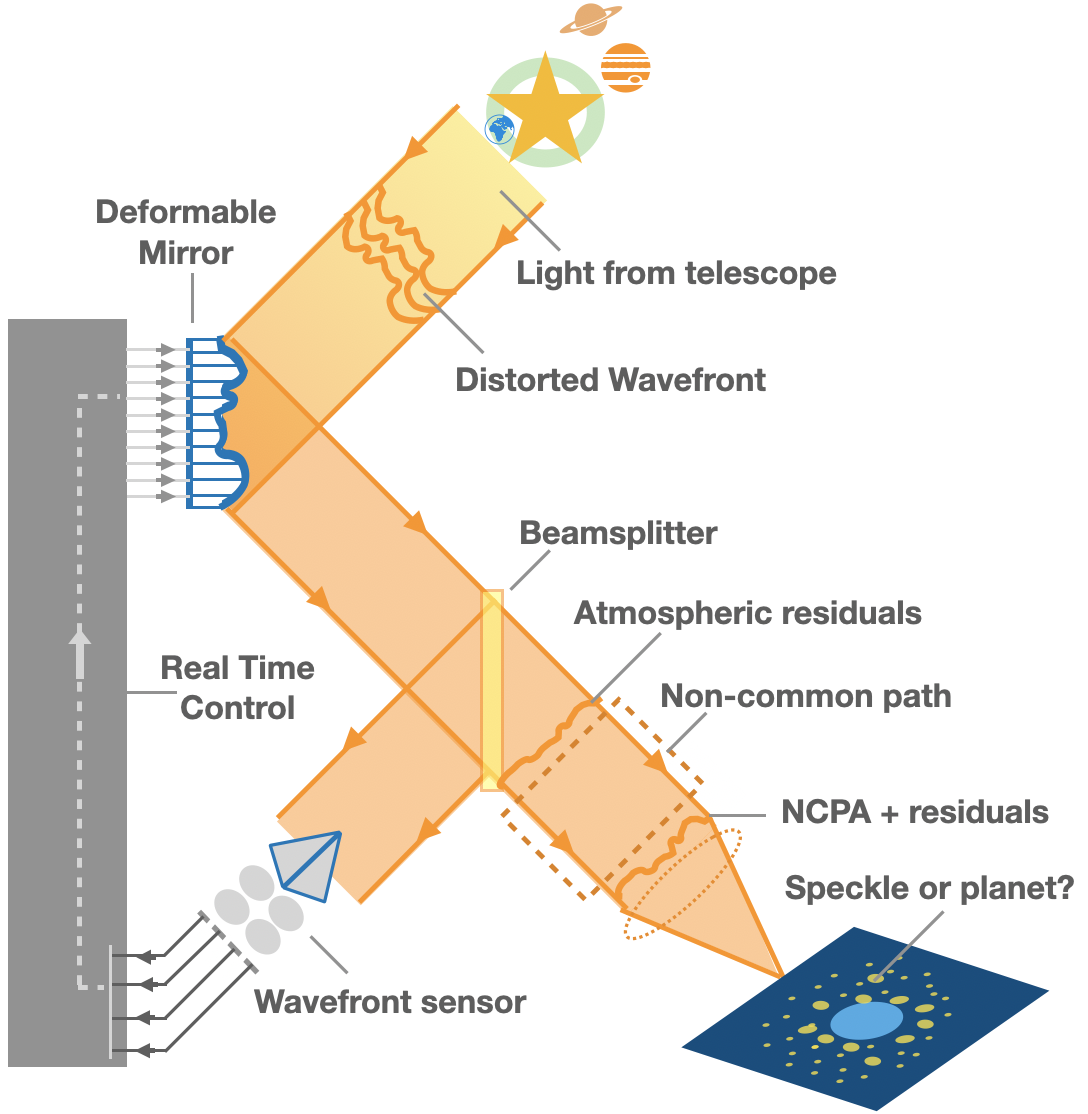}
\caption{Schematic view an AO system, presenting the problem of NCPAs, especially in the case of exoplanet imaging with a coronagraphic image.}
\label{fig:AO_schema}
\end{figure}

The ExAO loop corrects wavefront aberrations measured by the WFS, ideally approaching an aberration-free image. However, the ExAO loop convergence point, defined by the WFS reference, may not correspond to the optimal image quality. Reasons for such a discrepancy include:
\begin{itemize}
    \item{\textbf{Non-common path aberrations} (NCPAs) due to optics located after the beam splitting between WFS and science paths, inducing differential aberrations between the science camera and the WFS, as seen on Figure~\ref{fig:AO_schema}.

    In a high contrast imaging system, these aberrations may include coronagraph optics defects. These aberrations can also vary with temperature and mechanical deformation at timescales from minutes to hours, or with the positioning errors of moving optics, making them particularly challenging to calibrate. They are typically of the order of tens of nanometers, enough to lead to static and quasi static speckles in coronagraphic images \citep{sauvage2007, Vigan2019}. These speckles are also present in non coronagraphic images but less concerning because they do not contribute significantly to the SR, as usually beneath the camera contrast limit for unsaturated images}. 
    \item{\textbf{WFS calibration errors} due for example to a non-flat wavefront used to acquire a calibration.}
    \item{\textbf{Chromaticity} for systems where the wavelength of the WFS and the scientific camera are different.}
    \item{\textbf{Choice of optimal wavefront state}. The goal of the AO system may not be to produce a flat wavefront. For example, in a HCI, cancellation of static diffraction features (Airy diffraction rings, telescope spiders), such as dark hole techniques \citep{Potier2020darkhole}, may seek to drive the AO loop to a non-flat wavefront state.}
\end{itemize}

The phenomena responsible for the above effects have timescales slower than the millisecond-level atmospheric turbulence timescale, and can be (nearly) static. 
In this paper, we refer to all discrepancies between WFS reference and optimal image wavefront as NCPAs, noting that in previous publications it usually only refers to differential optical defects.

NCPAs affect all AO systems, but are particularly acute in high contrast imaging where small wavefront errors can easily mask faint exoplanet images (cf Figure \ref{fig:AO_schema}). Furthermore, quasi-static aberrations do not average out to a spatially smooth halo as fast as the atmospheric speckles, leaving structures in the coronagraphic image which resemble planets, unlike speckles generated by atmospheric turbulence alone.

The most reliable approach to measure and correct optical misalignments is to use focal plane images~\citep{Gonsalves-div-82}. In the particular case of NCPAs, these images are ideally generated by the science detector. 
Several solutions have been suggested for static NCPAs \citep{Paxman1988, Frazin2018,Vigan2018NCPA, vievardSPIEWFSscexao, BosFF2020, Potier2020darkhole, Codona_kenworthy}, the quasi static part being particularly challenging  to measure and correct. The evolutionary timescale of the aberrations with respect to the frequency of the correction is a main difficulty.

In a closed loop AO system, NCPAs are usually compensated by subtracting a biasing reference signal
in the WFS reference corresponding to the aberration to correct. Doing so requires good WFS response knowledge and stability, so that the adequate offset can be applied and maintained. Ideally, this offset should be updated in nearly real-time due to optical and mechanical variations in the instrument, however in practice it is typically updated on a daily or weekly basis. Finding the adequate offset requires focal plane images to update the reference, and can furthermore be particularly challenging if the WFS exhibits non-linear response that must be accounted for, as is the case in some high-performance WFSs such as the Pyramid WFS (PyWFS) \citep{Esposito2020,Deo2019gain,Chambouleyron2020, Chambouleyron2021}.

In this paper we will present the DrWHO algorithm, a model-free focal plane wavefront sensing approach which is aimed at finding the offset mentioned above to correct slow and static wavefront aberrations, including the NCPAs, on a few seconds timescale. After stating the issues to be solved in Section \ref{sec:problem_statement}, we describe the algorithm in Section \ref{sec:algo_des}. Numerical simulations with the COMPASS software \citep{Compass} are presented in Section \ref{sec:compass}. 
In Section \ref{sec:drwhoscexao} we show the results of the algorithm as obtained on sky with SCExAO.

Section \ref{sec:results} presents the main results of the on-sky validation, in terms of PSF and WFS reference evolution. In Section \ref{sec:Discussion}, we review the characteristics of the algorithm and describe further implications for use in HCI. 

\section{Problem statement: What is a good WFS reference?}\label{sec:problem_statement}

In a closed loop AO system, each WFS measurement is compared to a WFS reference (usually by  subtraction after flux normalization), to isolate residual wavefront errors that must be corrected by the DM. 

The real-time controller performs a wavefront reconstruction,
 
thus bringing the WFS signal to where it gives a flat wavefront, such that the WFS reference defines the convergence point of the AO control loop.

 The WFS reference can be measured with an internal light source inside the instrument, before on-sky observations. This can be done as a standalone step or by averaging the WFS signal when taking a response matrix, corresponding to the WFS signal when there is no aberration induced. However, in reality, the \textit{ideal} WFS reference may be different from the internal source WFS reference due to optical illumination discrepancies. The WFS reference is also not static, constantly evolving due to quasi-static aberrations and wavefront chromaticity. The accurate calibration of the WFS is therefore a significant challenge for any AO system.

A continuous way to measure the \textit{ideal} reference is therefore required for high accuracy wavefront correction. 

Finding this \textit{ideal} reference is particularly critical to ExAO systems, where a slight deviation can introduce slow/static speckles that appear similar to a planet image.\\

\section{Algorithm description}\label{sec:algo_des}

Since the goal is to ensure the AO loop converges to the \textit{ideal} reference, the \textit{actual} reference should be updated continuously. To do so, DrWHO regularly re-estimates this reference, chosen by selecting the WFS images corresponding to best PSF image quality. The algorithm is thus a lucky imaging selection of the PSF leading to the selection of the best WFS images. 

The algorithm first requires the focal plane PSFs and WFS frames streams to be synchronized in time, as described in Appendix \ref{appendix}. The algorithm furthermore requires that there exists a one-to-one mapping from PSF to WFS images, which is explored in section \ref{sec:mapping}.\\
The quality metric (\emph{score} hereafter) for the PSF is flexible as far as it optimizes the PSF quality (cf section \ref{sec:mapping}): it can be SR, contrast, minimum of intensity, etc. The algorithm proceeds as follows: on a defined timescale (set arbitrarily at 30 seconds), the algorithm first selects the best PSF frames according to a score metric and a predefined selection fraction (arbitrarily set to 10\%). Then, out of those 10\% frames, the corresponding WFS raw frames are extracted from the AO telemetry and averaged; the resulting WFS frame replaces the WFS \textit{actual} reference, and the algorithm is iterated to continuously optimize the WFS reference. \\

As the algorithm proceeds, the AO loop point of convergence removes NCPAs and other slow aberrations. Thus, the PSF quality improves, and converges towards a better score. 
The temporal frequency of the aberrations that DrWHO can correct depends on the timescale of one iteration of the algorithm (for a timescale of 30 seconds, the algorithm will be able to correct dynamic aberrations up to 0.03Hz).

We present the procedure step by step of DrWHO with the  description in Algorithm \ref{algo:DrWHO}.

\begin{algorithm}
\KwData{\begin{itemize}[noitemsep,nolistsep]
     \item WFS frames
     \item fast science camera frames
 \end{itemize}}
 \KwResult{Update the reference of the WFS}
 \textbf{Initialization:} take the reference of the WFS with the internal light source\;
 \While{AO loop is running}{
 \begin{itemize}
     \item acquire WFS and fast science focal plane images\;
     \item synchronize data cubes\;
    \While {DrWHO iteration (30 s) is running} {
    \begin{itemize}
        \item select the 10\% best science data\;
        \item select the corresponding WFS images\;
        \item average selected WFS images\;
        \item the resulting WFS frame replaces the WFS\\
        reference\;
        \item WFS reference updated. 
    \end{itemize}
   }
   \item restart DrWHO iteration
   \end{itemize}
   Change of AO point of convergence;
  }
\caption{DrWHO algorithm}
\label{algo:DrWHO}
\end{algorithm}

\section{Validation via numerical simulations}\label{sec:compass}

This section presents the AO simulation set up with the COMPASS simulator \citep{Compass}.
COMPASS is a versatile AO simulator, which has been designed to meet the need of high-performance for the simulation of AO systems, modeling several kinds of AO features, including several types of WFS, atmospheric simulations, DMs, telescopes, and RTCs. It offers a suitable environment for simulating Single Conjugate Adaptive Optics (SCAO), for 8-m class telescopes and 30-m class telescopes as described in \citet{Vidal2018}.  Simulations of DrWHO on the COMPASS software have first been implemented to explore the algorithm feasibility and potential.

\subsection{Extreme AO simulation}

The first step was to simulate an ExAO on COMPASS. We looked for simulating an AO close to the SCExAO instrument, which is a high-order AO system mounted behind Subaru's facility adaptive optics AO188 \citep{AO188_minowa} that provides a first level of correction (woofer) using a 188-actuator DM and a curvature WFS.

SCExAO performs high-contrast imaging thanks to a second level of correction (tweeter), using a 2000-actuator DM and a visible PyWFS.
We simulated a simplified ExAO tweeter system inspired by SCExAO, without trying to be an exact copy. Table \ref{tab:simuParams} synthetizes the parameters used in the COMPASS simulations.\\

\begin{table}[t]
		\centering
		\caption{%
			ExAO numerical simulations parameters.
		}
		\label{tab:simuParams}
		
		\renewcommand{\arraystretch}{1.2}
		\begin{tabular}{ll}
			\multicolumn{2}{c}{\textbf{Numerical simulation configuration}}\\
			\hline\hline
			\multirow{4}{*}{Telescope} & $D$ = 8.~m diameter\\
			& $\quad$No support spiders\\
			& Central Obstruction = 0.12 \\
			\hline
			\multirow{5}{*}{Turbulence layer} & von Kármán, ground layer only\\
			& $r_0$ = 0.16 at 500~nm \\ % variable in a\\
			%& $\quad$useful range of 7.0 - 35.0~cm\\
			& L$_0$ = 20~m \\
			& (to simulate post-woofer residuals) \\
			& $||\overrightarrow{\mathbf{v}}||$ = 10~m.s$^{-1}$\\
			\hline
			PyWFS & \\
			$\quad$Subapertures & 64$\times$64 \\%-- pixel size 42~cm.\\
			%& 24,080 useful pixels\\
			$\quad$Wavelength & Monochromatic, 750~nm\\
			%$\quad$Throughput & 50\% \\
			$\quad$Guide star magnitude & 7 \\
			$\quad$Modulation & Circular, 3~$\frac{\lambda}{D}$ radius, 24 points\\
			$\quad$Photon noise only \\% & 0.3~$e^{-}$\\
			\hline
			Deformable mirror & \\ 
			& 48 actuators across the diameter\\
			\hline
			Focal Plane Camera & \\
			$\quad$Wavelength & Monochromatic, 1.65~$\mu$m\\ 
			$\quad$Photon noise only \\
			\hline
			RTC controller & \\
			$\quad$Loop rate & 2~kHz\\
			$\quad$Method & Linear modal integrator\\
			%$\quad$Latency & Data flow 1 frame + MVM 1 frame\\
			$\quad$Basis & DM Karhunen-Loève (KL) basis\tablefootmark{(a)} \\
			$\quad$Loop gain & 0.3\\
			$\quad$Controlled modes & 1505\\
			$\quad$Modes filtered & 300\\
			\hline
		\end{tabular}
		\tablefoot{
			\tablefoottext{a}{- \cite{Compass}}
		}
	\end{table}

The simulation is idealized, with no source of noise other than only the photon noise on both the WFS and the science detector (observing mR=7 with 50\% efficiency), and does not take into consideration the telescope spiders. Furthermore, we considered a PyWFS with a higher sampling than on the SCExAO instrument (64 pixels over the PyWFS pupil diameter, versus 50 pixels), hence higher performance than achieved on-sky. The PyWFS is working in the visible at 750~nm, and the detector in the infrared at 1650~nm (cf Table \ref{tab:simuParams}). Finally, the COMPASS simulation allows to have synchronized exposure times, which is not the case on SCExAO (cf Appendix \ref{appendix}).

The quality criteria used for quantifying the PSF on COMPASS is the SR.

\subsection{DrWHO on COMPASS}
We implemented the DrWHO algorithm as described in Algorithm \ref{algo:DrWHO}. 
Results are presented for a case with 15 DrWHO iterations, each consisting of 10 000 AO loop iterations (corresponding to 5 seconds at 2kHz). The number of AO loop per DrWHO iteration was constrained by simulation time. 
The focus will remain on the low order modes.

\begin{figure}[t]
\begin{center}
\includegraphics[width=0.4\textwidth]{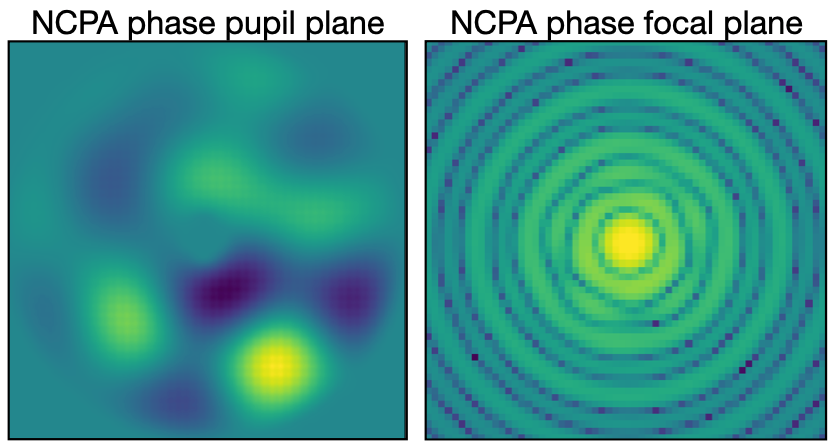}
\caption{Left: phase of the NCPAs in the pupil plane, with an amplitude of 30~nm RMS. Right: Corresponding PSF of size 60x60 pixels, or 12 $\lambda/D$, at 750~nm.}
\label{fig:NCPAs_psf}
\end{center}
\end{figure}

NCPAs, corresponding to a linear combination of the first 12 Karhunen-Loève modes (which is the linear and orthogonal modal basis used for computing the response matrix), with randomized amplitude, were applied on the science path (not visible to the WFS), with a total amplitude of 30~nm RMS (cf Figure \ref{fig:NCPAs_psf}). Such amplitude for the NCPAs is slightly greater than what is expected in reality. In fact, on SCExAO, NCPAs are expected to have a total amplitude of approximately 20~nm. Furthermore, in those simulations, the NCPAs only include differential optical aberrations between the science camera and WFS.

On SCExAO, and on most ExAO systems, the impact of NCPAs over the SR is relatively small, therefore it is not the best metric to quantify the efficiency of the algorithm, as the AO is dominated by dynamical wavefront residuals. In fact, NCPAs are usually small enough that they do not affect the SR significantly relative to dynamical wavefront residuals, but they add quasi-static speckles in the focal plane that can mimic planets, especially when a coronagraph is used. However, the SR is a good metric to make sure the algorithm does not diverge, in particular through the high-order modes as they could adopt a random walk behavior. Furthermore, the simulation presents a simplified system where a coronagraph is not simulated, hence the SR remains useful.

Table \ref{tab:compass_comp} presents the SR for different cases: when the loop is running without NCPAs, the SR reaches 87.9\% at 1.65 \textmu m. When NCPAs are added and left uncorrected, the SR is reduced to 86.8\%. Figure \ref{fig:NCPAs_psf} shows the impact of the NCPAs on a PSF free of other aberrations. 

\subsection{Simulation results}
\subsubsection{Results in terms of PSF quality}

\begin{table}[]
\begin{tabular}{l|ccc}
                               & \multicolumn{1}{l}{No NCPAs} & \multicolumn{1}{l}{with NCPAs} & \multicolumn{1}{l}{with DrWHO} \\ \hline
SR LE (\%)                          & 87.9                      & 86.8                         & 87.7                           \\
$\sigma_{OPD}$ (nm RMS) & 94.3                      & 98.8                         & 95.1                         
\end{tabular}
\caption{Comparison of the long exposure (LE) Strehl ratio between the following cases of the AO loop: no NCPAs, NCPAs alone, and post DrWHO with the NCPAs. The second line corresponds to the RMS of the wavefront for each of those cases, calculated from the SR of the line above with the Maréchal approximation.}
\label{tab:compass_comp}
\end{table}

\begin{figure}[t]
\begin{center}
\includegraphics[width=0.52\textwidth]{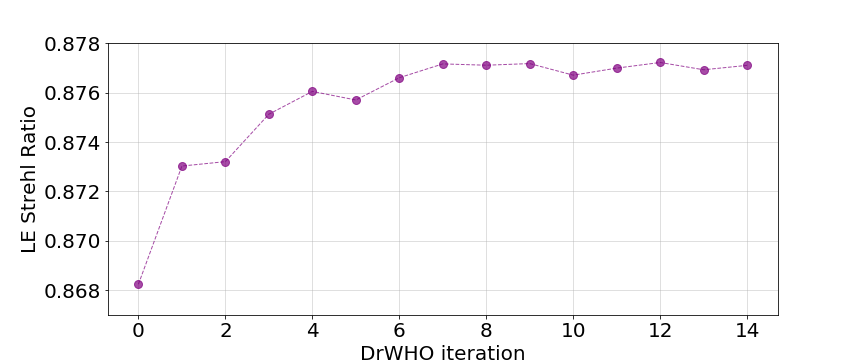}
\caption{Evolution of the long exposure (LE) Strehl ratio during the DrWHO run on COMPASS, corresponding to 15 DrWHO iterations of 10 000 AO loop each.}
\label{fig:compass_LE_SR}
\end{center}
\end{figure}

Figure \ref{fig:compass_LE_SR} and Table \ref{tab:compass_comp} present the result of the SR after the DrWHO run: the final SR after the 15 iteration DrWHO run reaches 87.7\%, with a maximum of 87.9\% when there are no NCPAs. As shown on Figure \ref{fig:compass_LE_SR}, nearly half of the SR improvement is obtained within the first iteration of DrWHO. Then, the SR keeps slowly improving, until it reaches a plateau. After 15 iterations, the SR reached is almost the same as the SR measured without NCPAs. \\

In order to better understand the impact on the optical path difference (OPD), we used the Maréchal approximation

\begin{equation}
    \text{SR} = \exp(-\sigma_{\Phi}^2) = \exp\left(%
    -\dfrac{4\pi^2 \sigma_{\text{OPD}}^2}{\lambda^2}
    \right),
\end{equation}
with $\sigma_{\Phi}$ being the variance of the phase, and $\sigma_{\text{OPD}}$ being the variance of the OPD.

The corresponding values for $\sigma_{OPD}$, calculated from the measured SR, can be found in Table \ref{tab:compass_comp}. 

The wavefront residuals are dominated by dynamical atmospheric residuals, with 94~nm RMS in the ideal case without NCPAs. The quadratic diffrence with and without NCPAs is 30~nm RMS of NCPAS, matching the amplitude of the modes added in the simulation. Of these 30~nm, 26.6~nm (89\%) are corrected by the DrWHO algorithm. 
\\

\subsubsection{Results in terms of modes corrected}

\begin{figure}[t]
\begin{center}
\includegraphics[width=0.45\textwidth]{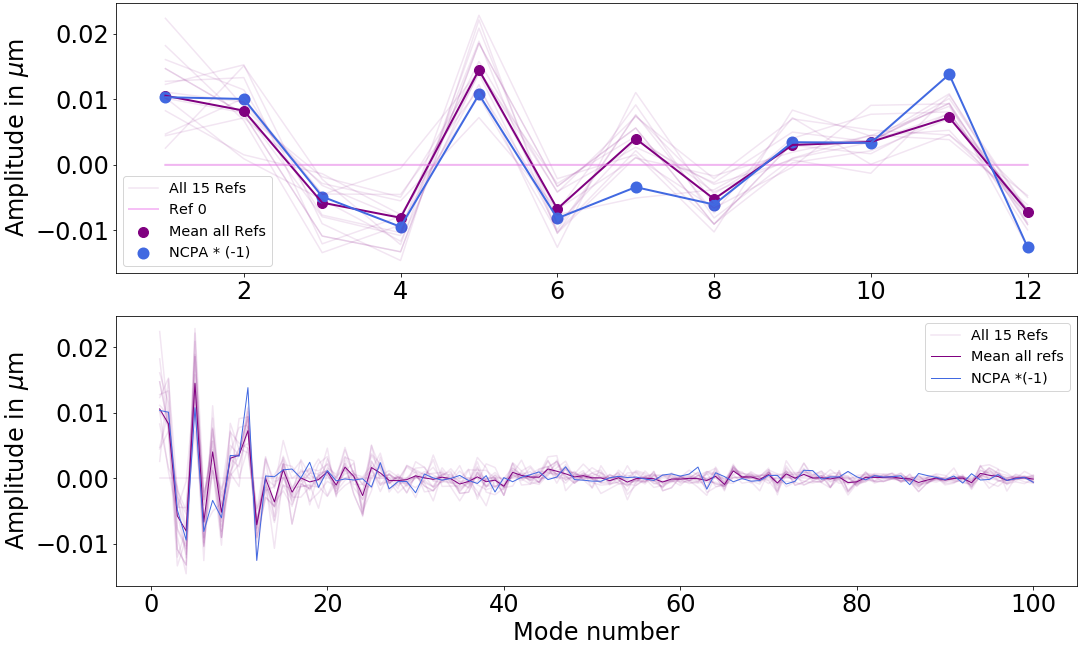}
\caption{Top: projection of the references provided by the 15-iteration DrWHO run over the modal basis, with the projection of the opposite of the NCPAs, in order to better visualise the correction. Bottom: projection over the first 100 modes.}
\label{fig:ho_modes}
\end{center}
\end{figure}

The 15 references for each iteration of the DrWHO run were projected over the modal basis with which the response matrix was acquired. Figure \ref{fig:ho_modes} shows this modal decomposition along with the ideal NCPA correction, where the opposite of the NCPAs was added, to better compare the correction. It appears from the top plot of this figure that, as we previously concluded, the algorithm converges from the first iteration towards a compensation of the NCPAs. The average of all those references (minus the reference 0, which is the initial reference) is neighbouring the NCPAs aberrations. This proves that DrWHO measured them, with the correct amplitude, and compensated for them. 
The middle and bottom plots of this figure prove that the algorithm does not diverge at the higher order modes, as confirmed by the increase of SR. However, they add a noise contribution to the correction.
The quadratic sum of 12 modes from the mean of all the references from the algorithm corresponds to \textbf{36~nm} RMS, which is roughly consistent with the amplitude of the NCPAs.

\subsubsection{Discussion}
DrWHO on COMPASS improved the image quality in terms of SR, and the NCPAs applied over the PSF, unseen by the WFS, were almost totally corrected. Half of the correction has occurred at the first iteration of the algorithm. The projection of the references from DrWHO is showing that the NCPAs are nearly entirely compensated for, and the higher orders, where we did not apply NCPAs, are not diverging. 
However close the SR got to the SR without NCPAs, it did not reach perfectly the original pre-NCPAs SR. Besides the fact that SRs are defined up to a certain precision only \citep{Roberts2004Strehl}, this can be explained by two effects: first, as seen in Figure \ref{fig:ho_modes}, the modes are not perfectly compensated; and second, the modes of the references beyond the ones applied on the NCPAs (12 modes) did have a small -yet not fully negligible- impact on the reference update, hence they may have slightly impacted the SR.

\section{DrWHO on SCExAO}\label{sec:drwhoscexao}

The DrWHO algorithm was then adapted and deployed on the SCExAO instrument, which is located on the infrared Nasmyth platform of the Subaru Telescope, downstream of the AO188 instrument.

Several aspects are different between the COMPASS simulation and the SCExAO instrument, however the concept remains the same.

The instrument is equipped with a Pyramid WFS operating in the 600-950~nm wavelength range \citep{Lozi2019PWFS}. The real time control (RTC) of the AO system is managed by the CACAO software - Compute And Control for Adaptive Optics - \citep{cacao2018}. The DM has 45 actuators across the pupil, giving a control radius of 22.5$\lambda$/D. We use the CACAO software to interact with the system and implement the algorithm, to communicate between the PyWFS and the DM.
%CACAO provides 12 different DM control channels, on which several corrections are implemented from additional wavefront sensors or wavefront control techniques \citep{BosFF2020}.
%Furthermore, CACAO updates the reference of the PyWFS by adding an offset which prevents the AO loop to cancel out the commands of the wavefront control sensing techniques with the DM. 
There are several scientific modules downstream of SCExAO. The one used for this paper was one of the two VAMPIRES camera, which is an instrument operating in the visible \citep{vampires2015}, with a pixel scale of 6.1~mas.

On SCExAO, the PyWFS and the short exposure focal plane camera run at different frequencies, with typical frequencies of 2kHz for the former, and between 200 Hz to 2kHz for the latter, hence the need to synchronize the data streams to run DrWHO. This is further described in Appendix \ref{appendix}.

\subsection{PSF quality Flux criteria}

We introduce the Flux Concentration (FC), which is the $\alpha$-norm defined as 
\begin{equation}
\centering
    \text{FC} = \frac{\sum_i  (x_i^{\alpha})}{ (\sum_i x_i ) ^{\alpha}},
    \label{eq:FC}
\end{equation}

with $x_i$ being the value of the pixel i, and $\alpha > 1$. If some pixels have a negative value following a dark subtraction due to camera readout noise, then the pixel value is set to zero. 
This quantity is independent of absolute flux, since it is normalized by the total flux as seen in the denominator of equation \ref{eq:FC}. 

The FC approximately tracks SR, but is less sensitive to noise and sampling effects, as it considers an ensemble of pixel values instead of the PSF peak.  If the intensity is concentrated in one pixel, FC equals 1. For the intensity spread over several pixels, FC < 1. If the light is equally spread over many pixels, then FC tends to 0. We will be using $\alpha = 2$, noting that higher values of $\alpha$ favors the brightest pixels. \\

\begin{figure}[t]
\centering
\includegraphics[width=0.32\textwidth]{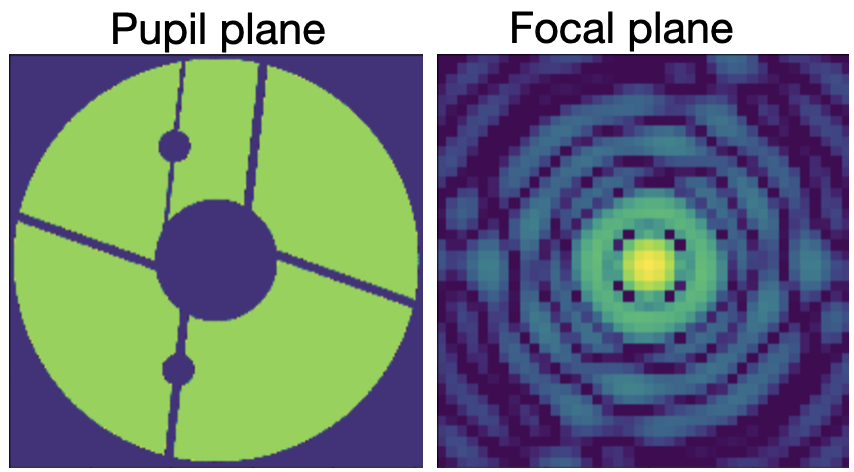}
\caption{Left: simulated pupil plane, including the spiders and a mask for two dead actuators. Right: simulated perfect SCExAO PSF}
\label{fig:perfect_psf}
\end{figure}

We choose to compute FC for a subarray of the full image covering the central region of the PSF, in order to focus on the low order modes: wavefront aberrations at low spatial frequencies (except for Tip-Tilt) redistribute the flux in such a way that it spreads across more pixels in the central region of the PSF, and thus the higher the low order aberrations will be, the lower the FC will get. 
The FC can be normalised by the FC of a simulated perfect PSF---which we will call 
FC$_0$ - 
computed for the pupil of the telescope (cf Figure \ref{fig:perfect_psf}), as the FC is a relative number. 
This new quantity is referred to as the normalized FC (nFC): nFC = FC / FC$_0$.

\section{Testing the principle on sky}\label{sec:results}

\subsection{Observation setup}

We performed observations on the engineering night of December 8 UT, 2020, during which we ran DrWHO over a period of 10 minutes, corresponding to 20 iterations of 30 seconds. Each iteration entails synchronized data cubes resampled to a common 1 kHz frame rate for both VAMPIRES and PyWFS images, running at 500Hz and 2kHz respectively. 1200 modes were corrected. The observation conditions at that time were a seeing of 0.5", and a wind speed of 4 m/s. The observed target was the star $\xi$ Leo star, at an airmass of 1.03. The window size of the image for the selection of the quality criteria was of 70x70 pixels, corresponding to 0.43x0.43 arcsec field of view, and to approximately 250 modes. 

\subsection{Evolution of the PSF quality}
\begin{figure}[t]
\includegraphics[width=0.45\textwidth]{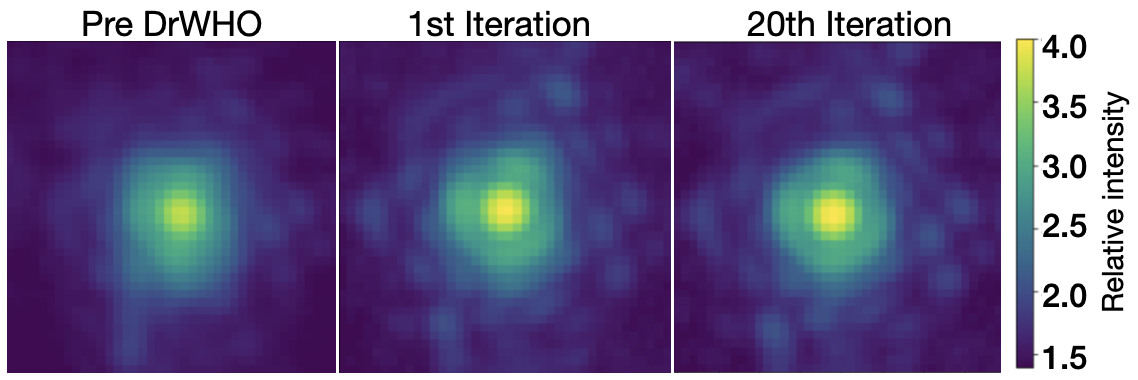}
\caption{Evolution of the on-sky PSF before running the algorithm, after the first iteration, and the after last iteration. Each image is 0.25~arcsec (40x40 pixels) across, acquired at $\lambda = $ 750~nm, 30 sec exposure time (computed by co-addition of 15,000 frames acquired at 500~Hz). Those PSFs should be compared to the ideal PSF in Figure \ref{fig:perfect_psf}, with matching wavelength, field of view and orientation of Figure \ref{fig:SkyResult} shown in logarithmic brightness scale.}
\label{fig:SkyResult}
\end{figure}

\begin{figure}[t]
\centering
\includegraphics[width=.47\textwidth]{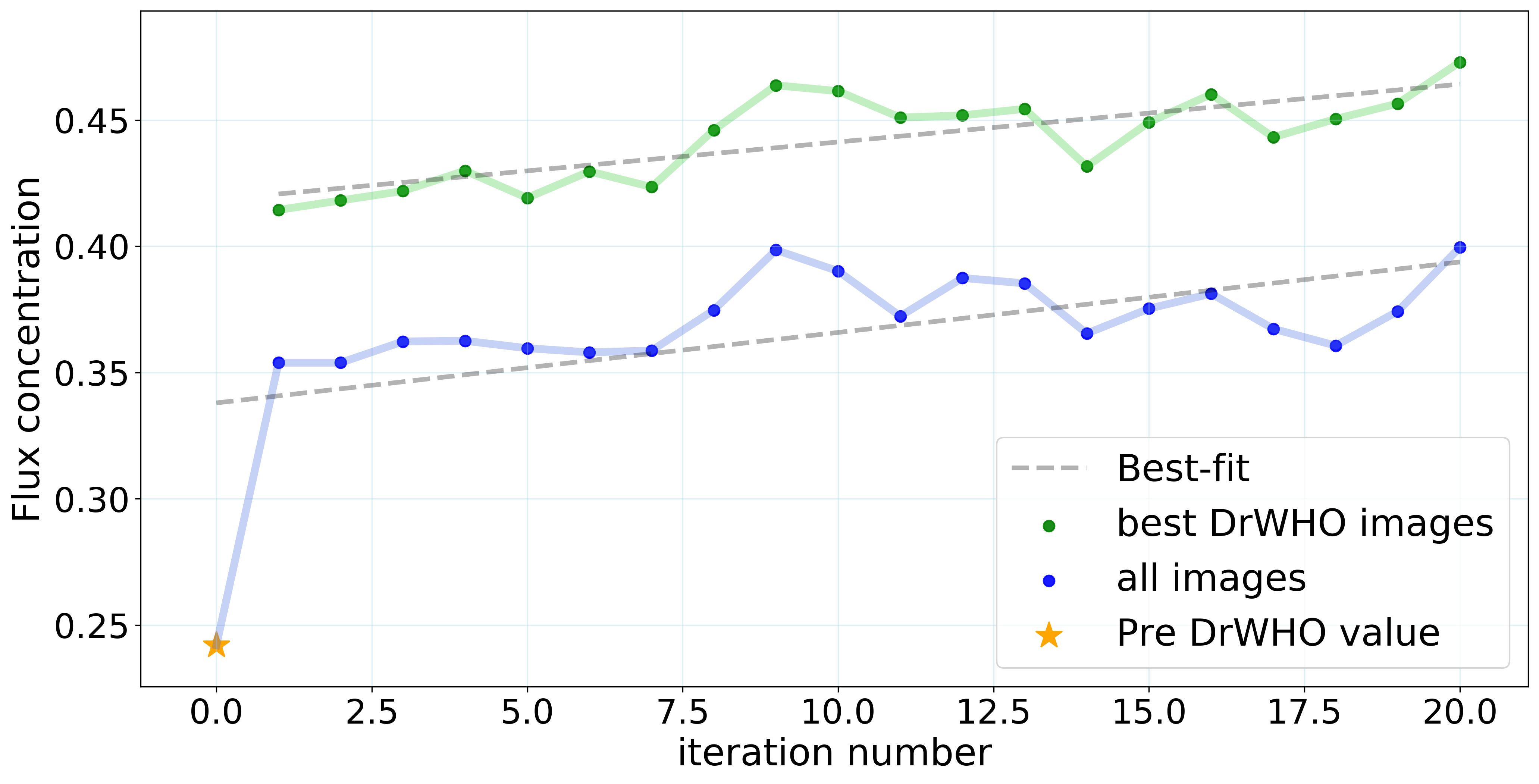}
\caption{Evolution of the FC over 20 iterations of DrWHO, corresponding to a period of 10 minutes, including the nFC of the PSF preceding the run. 
The blue curve corresponds to the evolution of the nFC of all the PSF averaged over the DrWHO iteration, while the green one corresponds to the evolution of the nFC of the 10\% PSF chosen by DrWHO. The best linear fit is presented. }
\label{fig:NNevolion}
\bigbreak
\includegraphics[width=.47\textwidth]{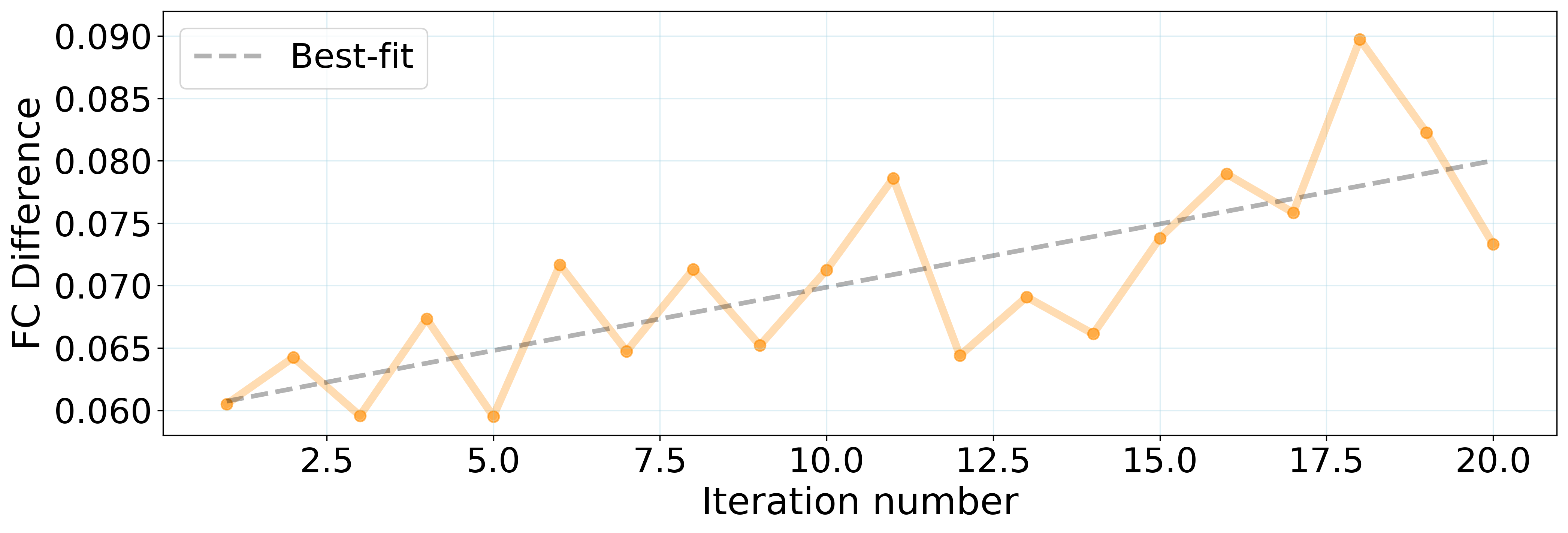}
\caption{Difference between the nFC of the mean of the best 10\% images, and the nFC of the mean of all the PSFs, over 20 iterations of DrWHO. We calculated the best linear fit to better understand the increase of the difference.}\label{fig:difference}
\end{figure}

\begin{figure}[t]
\centering
\includegraphics[width=.52\textwidth]{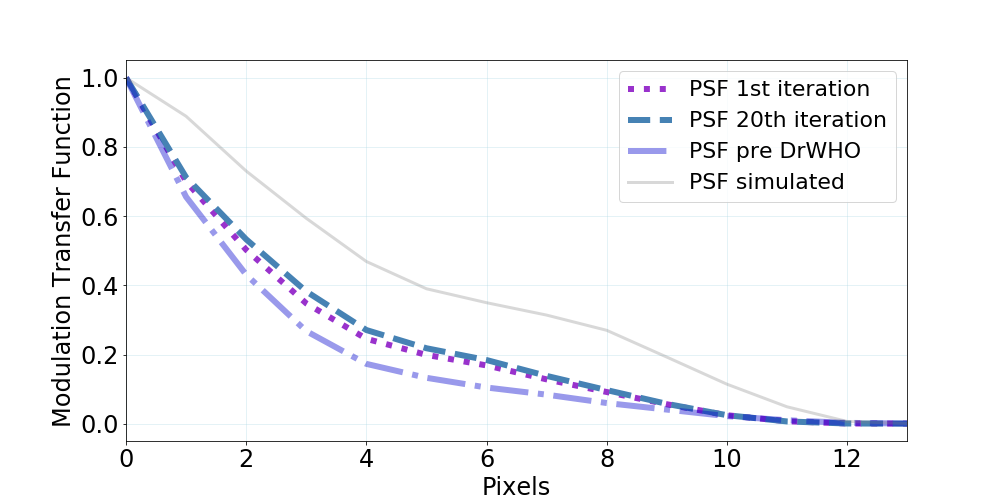}
\caption{MTF of the PSF before running DrWHO, of the 1st iteration, and the last iteration (cf Figure \ref{fig:SkyResult}), compared to the ideal simulated Vampires PSF.}
\label{fig:MTF}
\end{figure}

Figure \ref{fig:SkyResult} shows the PSF just before running DrWHO, followed by the PSF after the first iteration and the PSF after the last iteration of this run. After 20 iterations, the PSF visually looks better and more circular, attesting a partial correction of the low orders. Furthermore, the outer part of the PSF looks darker compared to the first iteration. The improvement in PSF quality is stronger between the pre-DrWHO PSF and the first iteration PSF, compared to the first iteration PSF and the last one. Hence, it appears that a major part of the correction was applied by the first iteration, like in simulations. \\
On the resulting PSF of size 70x70 pixels for each iteration, we calculated the nFC over a smaller window of 40x40 pixels (0.25 arcsec field of view), focused on the central part of the PSF. The images were background subtracted and re-centered. 
Figure \ref{fig:NNevolion} shows the evolution of these nFC over the 20 DrWHO iterations. We plotted the evolution of the FC for the 10\% best PSFs that were selected by the algorithm for the new reference computation, as well as the FC for the average of all the PSFs for each iterations. The difference of those nFC is shown in Figure \ref{fig:difference}.\\
Furthermore, the Modulation Transfer Function (MTF) has been calculated for the PSFs before and after the DrWHO run, and is shown in Figure \ref{fig:MTF}. The MTF of the ideal simulated PSF has been added to the figure. This figure shows as well that most of the correction is done at the first iteration.

Several points are worth noting: 
\begin{itemize}
    \item The major correction is applied at the first iteration, with the nFC jumping from 24.3\% to 35\%, comparably to simulations; 
    \item The FC keeps increasing throughout the run, as shown by the best linear fit of the two curves, which is something that was not observed in simulations, where the evolution was smoother and reached a plateau ;
    \item Considering PSFs averaged over each iteration, nFC increases from 28\%, to 35\% at the first iteration, to then 40\%: \textbf{the overall improvement is 15.7\%} ; 
    \item The best DrWHO images present a sharper increase than all the images together : Figure \ref{fig:difference} presents the difference of these two nFC over the algorithm run. Despite selecting only the best 10\% frames, DrWHO seems to be improving both the selected and unselected frames.

    We note that the algorithm has not fully converged yet and would have performed even better on a longer period of time ; 
    \item Finally, Figures \ref{fig:NNevolion} and \ref{fig:difference} show that even if the PSFs are subject to the variations in seeing over a few minutes, they overall get better in terms of the concentration of flux. Hence, DrWHO corrects the NCPAs even if they are not as dominant as the dynamical wavefront residuals from atmospheric turbulence.
\end{itemize}

\subsection{Evolution of the WFS reference}

\begin{figure}[t]
\centering

\includegraphics[width=.48\textwidth]{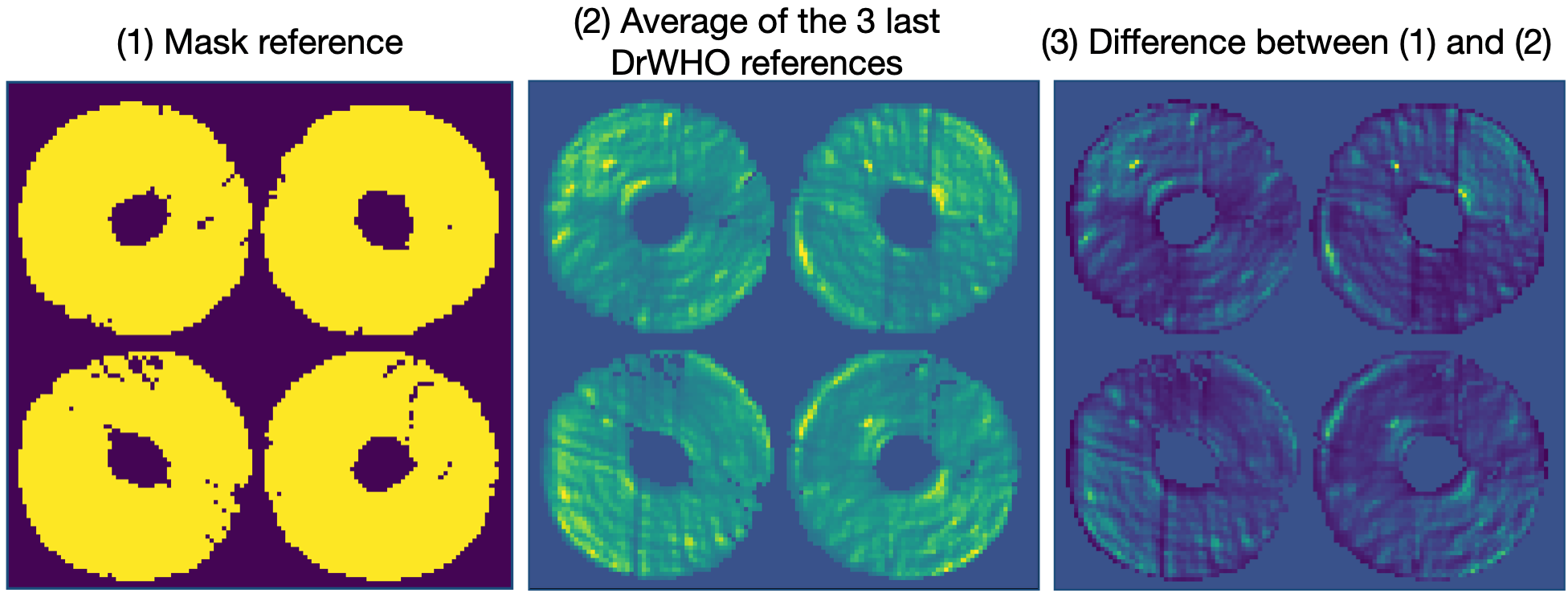}
\caption{Left: images of the PyWFS reference before the first iteration, referred to as \textit{Ref0}, set equal to the WFS mask. Middle: average of the last 3 iterations of DrWHO WFS references. The contrast scale for this image has been modified for better visualisation. Right: difference between the two images in the left and in the middle.}
\label{fig:reference_diff}
\end{figure}

\begin{figure}[t]
    \centering
    \includegraphics[width=.5\textwidth]{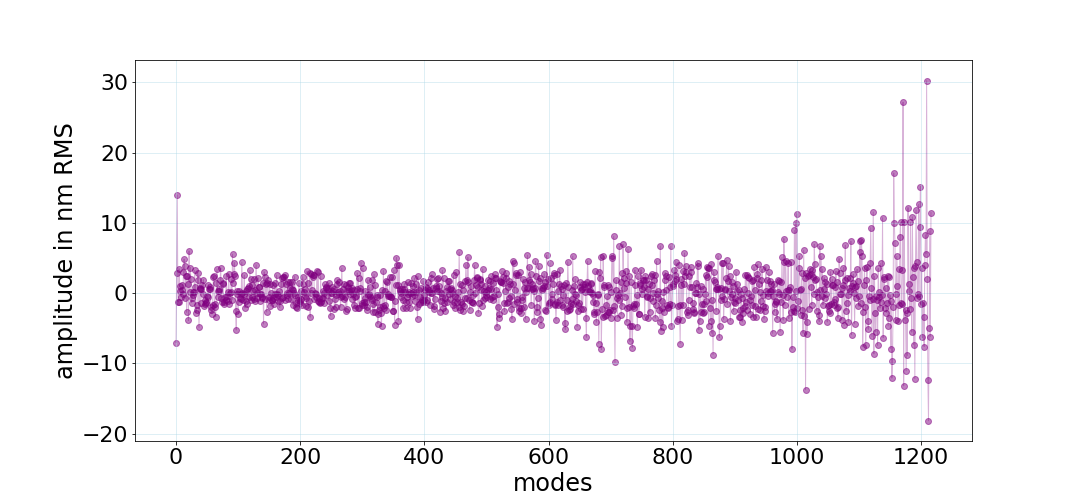}
    \caption{Modal decomposition of DrWHO correction, computed from the difference between the pre-DrWHO WFS reference and the average of the WFS references for the last 3 DrWHO iterations.}
    \label{fig:modes_proj}
    \bigbreak
    \includegraphics[width=.5\textwidth]{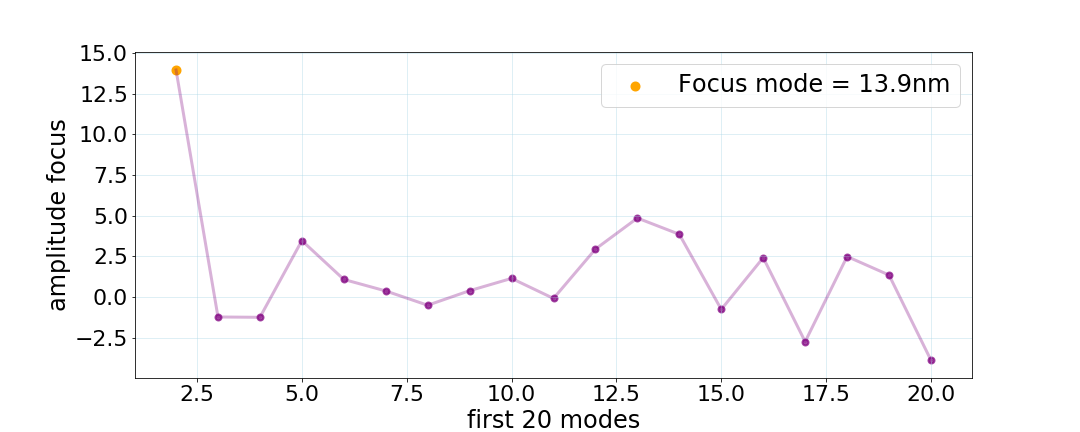}\\
    \includegraphics[width=.5\textwidth]{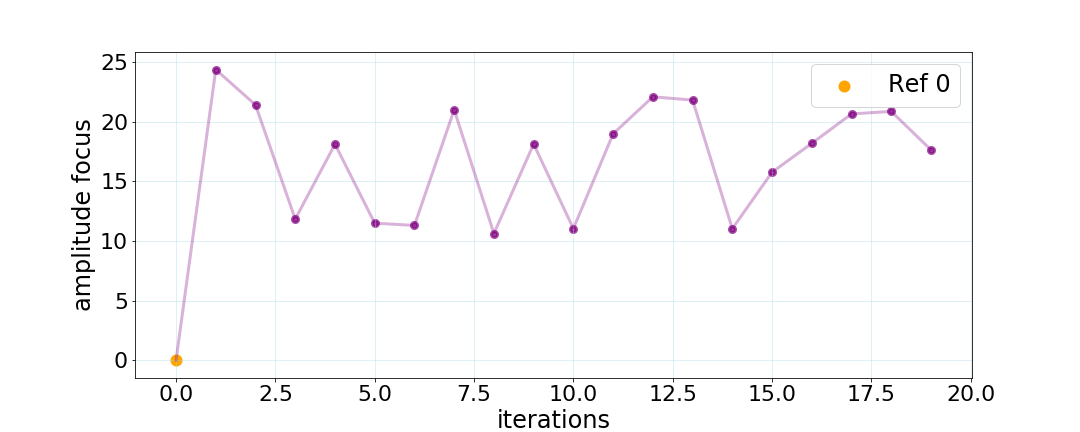}\\

    \caption{Top: modal decomposition of DrWHO correction for the first 20 modes, excluding tip-tilt. Focus is the major contributor with an amplitude of 13.9~nm RMS. 
    Bottom: evolution of the focus WFS reference correction during the DrWHO run. }

    \label{fig:10_modes}

\end{figure}

Figure \ref{fig:reference_diff} compares the WFS reference prior to the DrWHO run, Ref 0, which is simply the WFS mask showing which pixels are active (i.e. respond to DM pokes) in the WFS image, and the average of the last 3 WFS references applied by DrWHO, during the on-sky 10-minute run. The averaging minimises the noise contribution.
The difference between the two references is also shown in Figure \ref{fig:reference_diff}.
This difference was then converted to wavefront mode coefficients by multiplication by the control matrix. Modal coefficients are shown in Figure \ref{fig:modes_proj} where the first three modes are tip, tilt, and focus and the following modes correspond to orthogonal modes, optimized for SCExAO's AO control law.

The total contribution of all modes is measured to be 116~nm RMS. However, as seen on Figure \ref{fig:modes_proj}, the high order modes, corresponding to the higher spatial frequencies, seem to be noise more than actual correction. As described in section \ref{sec:mapping}, it is expected that a single metric cannot be sufficient to correct all 1217 modes.

Hence, we will focus on the low order modes, from mode 2 to 20. When calculating the total contribution of those modes, we measure a correction of 17.2~nm RMS.

Figure \ref{fig:10_modes} (top) shows the first 20 wavefront mode values, excluding tip-tilt. Focus is the dominant contribution to the aberrations, with an amplitude of nearly 14~nm, while the other modes have values below 5~nm.

We furthermore note that the higher modes, corresponding to higher spatial frequencies, have a non-negligeable contribution in the overall correction. The way the selection is done is less efficient on high dimensions.
DrWHO optimizes the PSF with over a thousand wavefront modes using a single scalar metric, here the FC. The low orders modes  are improved, however the high orders follow a random walk. There is a global selection, but that creates noise on the higher order modes, as can be seen in Figure \ref{fig:modes_proj}. This is further developed in section \ref{sec:Discussion}.

Table \ref{tab:mode_contribution} presents the contribution of the modes in nm RMS, in the following cases: in considering all the modes except tip-tilt (from 2 to 1217) ; then in removing the higher order modes (from 2 to 900 only) to show what the contribution would be without the higher spatial frequencies ; then the low order modes (from 2 to 20), and finally the focus contribution, which is  13.9~nm.

\begin{table}[ht]
\begin{tabular}{c|l|l|l|l}
\textbf{Modes}                                                            & 2 to 1217 & 2 to 900 & 2 to 20 & \multicolumn{1}{c}{Focus} \\ \hline
\textbf{\begin{tabular}[c]{@{}c@{}}Contribution \\ (nm RMS)\end{tabular}} & 116.3     & 70.3     & 17.2    & 13.9   
\end{tabular}
\caption{Contribution of the modes of the difference of Figure \ref{fig:difference} in nm RMS}
\label{tab:mode_contribution}
\end{table}

\section{Mapping between WFS and PSF}\label{sec:mapping}

 This section further explores the mapping between PSF and WFS images, in both directions, from PSF to WFS images and to WFS images to PSF. 
\begin{itemize}
    \item From WFS to PSF: we expect the PSF to be uniquely defined by the WFS image, as the WFS is designed to unambiguously measure the incoming wavefront phase, and the PSF is the square modulus of the wavefront complex amplitude. We note that the Pyramid WFS raw images also encode the wavefront amplitude, which may also affect the focal plane image. 
    \item From PSF to WFS images: in general, the WFS image cannot be unambiguously derived from the focal plane image. For example, the wavefront focus mode exhibits a sign degeneracy, as positive and negative focus values produce the same focal plane image. 
\end{itemize}

 This latter aspect is problematic for DrWHO, as the algorithm requires a one-to-one mapping from PSF to WFS images, since we select WFS images based on PSFs. Specifically, if two different wavefront maps produce the goal PSF image, DrWHO would converge to an average of the two wavefront maps. For example, a defocused PSF goal would be met by positive or negative focus values, but the average wavefront map produced by DrWHO would have both WF solutions cancel each other, yielding a zero=defocus solution that does not match the goal defocused PSF.
Thankfully, we expect this degeneracy disappears if the goal image is the aberration-free PSF corresponding to a flat wavefront, as only a flat wavefront can produce this PSF.  Hence, as DrWHO tries to optimise the PSF flux concentration, there should be a one-to-one mapping from PSF to WFS images. \\

 To test this point, we have analyzed the data cubes of synchronized frames from the observation night of section \ref{sec:results}, and used the approach described in \cite{Guyon_cluster_SPIE} to identify image pairs that are similar, as measured by the euclidean distance between images. 

\begin{figure}[t]
    \centering
    \includegraphics[width=.5\textwidth]{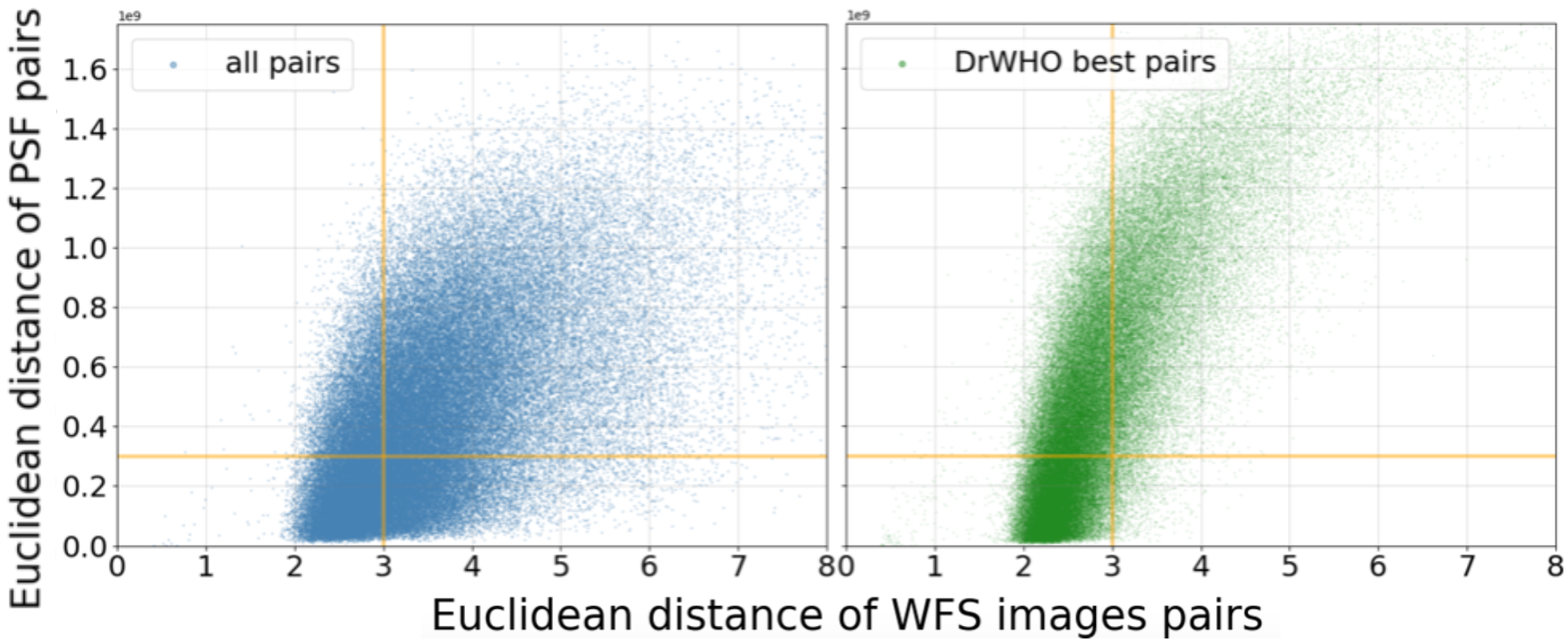}
    %\caption{(a)}
    \caption{ Each point of that figure corresponds to a pair of two epochs, to which we can associate a pair of images, both in WFS space and PSF space. The left hand side corresponds to all the pairs, without any selection; the right hand side corresponds to the 10\% best PSF selected by DrWHO.} 
    \label{fig:euc_dis}
    \bigbreak

    \includegraphics[width=.5\textwidth]{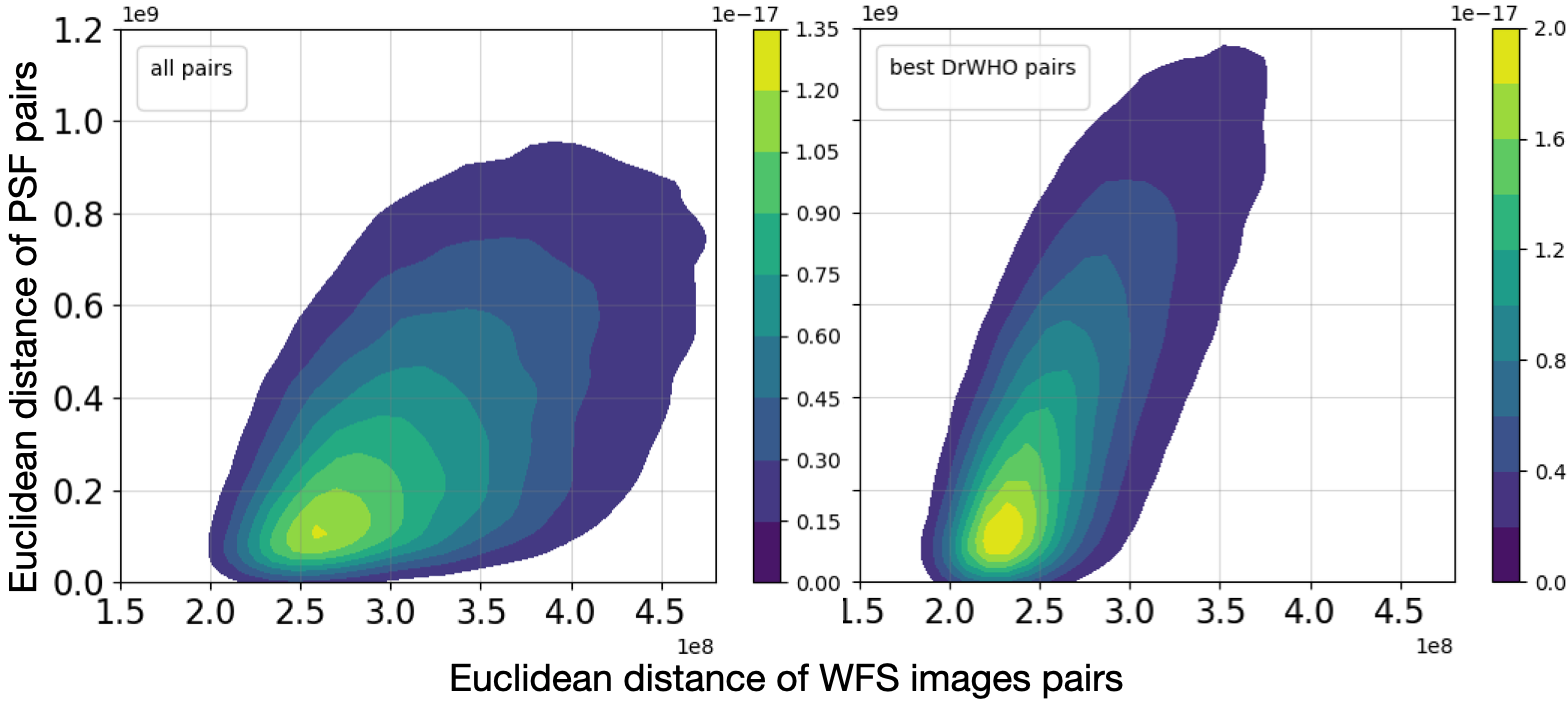}\\
    \caption{ Density plot corresponding to figure \ref{fig:euc_dis}. As seen with the colobar, the density plot of the best pairs selected by DrWHO is essentially focused near the origin, where the best PSF are located.} 

    \label{fig:density_plot}
\end{figure}

The result is shown in figure \ref{fig:euc_dis}. 
Each point maps to a set of two epochs, from which we extract the corresponding pair of WFS images, and the pair of PSF images. 
The y axis is the euclidean distance between the two PSFs, and the x axis is the euclidean distance between the two WFS images. Points in the lower part of the graph therefore correspond to epoch-pairs for which the focal plane PSFs are similar, while the left part corresponds to epoch pairs for which the WFS images are similar. 
Because the number of pairs is very high, we adopted the approach of looking at a random sub-set of 100,000 points out of 400 million. The corresponding density plot is shown in figure \ref{fig:density_plot}.

Horizontal and vertical lines have been added for visualisation help in figure \ref{fig:euc_dis}, both located at the same arbitrary value on each side. The horizontal line stakes out where the pairs of PSF are similar, and likewise for the vertical line and the pairs of WFS images. 
To validate the mapping from PSF to WFS images, we consider at the region below the horizontal line.
As seen on the figures \ref{fig:euc_dis} and \ref{fig:density_plot} on the left, for pairs of PSFs that are very close to each other (hence the part below the horizontal line) there are several pairs of WFS images possible (see the right part of the vertical line). This confirms our expectation that DrWHO should not work when not selecting for the sharpest PSF images (hence, it would fail when trying to maintain a defocused PSF for phase diversity, for example). However, on the Figure on the right, the fraction of data points on the bottom right part of the cross is almost zero. Points are still scattered due to numerous sources of noise, and some points still fall to the right of the vertical line. However, as seen in figure \ref{fig:density_plot}, the density of points is much higher in the lower left part of the distribution, indicative of a one-to-one mapping from PSF to WFS when selecting for the best PSF. This confirms the validity of DrWHO. \\

\section{Discussion}\label{sec:Discussion}

\subsection{The problem of dimensionality}
With DrWHO, two competing processes are at play: improvement by removal of dominant WF aberration modes, and random walk due to noise in the wavefront sensor and focal plane image. Our data demonstrate that DrWHO is able to improve WF quality over low-order modes, where relatively large errors are concentrated over a small number of modes. At high spatial frequencies, our current implementation is inefficient and likely introduces wavefront noise due to the random walk process. The challenge is ultimately linked to the curse of dimensionality, where optimization in a high dimension space with a single scalar metric is inefficient and unstable: the probability of all WF modes being simultaneously close to zero (no aberration) becomes vanishingly small, so the DrWHO solution follows a random walk around the optimal solution. To address this limitation, a multi-dimensional optimization metric should be adopted so that the optimization can be performed as parallel independent low-dimension optimizations. A natural implementation would be to separate the focal plane image in spatial zones, each corresponding to a set of wavefront modes according to the Fourier transform relationship between focal plane position and pupil plane spatial frequency.

\subsection{DrWHO characteristics}
Below are some noticeable characteristics of the DrWHO algorithm that have been demonstrated in this paper: 
\begin{enumerate}
    \item It is robust: the algorithm slowly but surely converges towards a better compensation of static and slow aberrations. As it is based on lucky imaging, it relies on some realisations of residual atmospheric turbulence to reduce wavefront aberrations. DrWHO contributes to getting closer to an absolute sensor, driving the WFS reference to optimize PSF in the focal plane.
    \item It does not rely on any model, and does not rely on using the DM for probing, making it compatible with other wavefront control techniques: DrWHO's approach is different from active cancellation speckle, because it simply does a statistical selection using the natural dynamical atmospheric residuals for probing, instead of adding artificial probes.
    \item It does not require modification of the AO loop control scheme other than WFS reference updates.
    \item It does not make any linear assumption, so it is not concerned with non-linearities of the system, such as WFS non-linearity.
    \item It is flexible in the parameters, whether it is the score (Strehl, contrast, intensity, etc.), or the selection fraction and the time of the iteration, making it adaptable to different weather conditions and AO systems with different characteristics and scientific goals.
    \item However, DrWHO needs a fast focal plane camera as close as possible to the science camera, to more accurately compensate the NCPAs and low aberrations. 
    \item The selection criteria for the PSF selection has to optimize the PSF, DrWHO will not work on a convergence point that is not the flat wavefront due to the fact that the mapping from PSF to WFS images cannot be established otherwise.
    
\end{enumerate}

On-sky tests were run on a bright star, but the algorithm efficiency must be improved to run on fainter targets, and be able to optimise deep contrast for HCI. Such application to HCI with coronagraphic images providing some adaptations of the quality criteria of selection, where it would be exactly the same thing as what is presented here, but with adapting the quality criteria of the selection. In this case, instead of selecting PSFs with the maximum FC, image contrast would be optimized. A first attempt at HCI application was presented in \cite{spie_drwho}.

\section{Conclusion}

We implemented the DrWHO algorithm first using an ExAO simulator, then validating on-sky using the SCExAO instrument at the Subaru Telescope.
Results presented in this paper demonstrate its ability to measure and compensate for NCPAs, especially the low order modes, considered to be a limiting factor in the detection and characterisation of exoplanets in high-contrast imaging observations. To quantify the quality of the PSF, we used the Strehl ratio for the simulations, and the \textit{Flux Concentration} for the SCExAO run. 
For the simulation, we observed a nearly perfect compensation of the added NCPAs, at the first iteration.

For the on-sky test, we combined the PyWFS measurements and the focal plane images in visible from the VAMPIRES module. 
When DrWHO was running, the PSF significantly improved (15.7\% relative improvement in flux concentration) over 20 iterations of 30 seconds, for a total of 10 minutes. The best PSF have a sharper increase over the run of the algorithm, reaching 48\% of the simulated ideal PSF flux concentration. 
The visible improvement in the image quality was confirmed with the calculation of the MTF before and after the DrWHO run.
The data on this run were used to explore the mapping from PSF to WFS images, as a one-to-one mapping is a necessary condition for DrWHO to converge to the goal PSF image. We established the uniqueness of the mapping in the case where the selection metric is optimal for the aberration-free PSF. \\
This shows that DrWHO is able to improve the wavefront quality arriving on the science camera, and partially correct for the static and quasi-static NCPAs, over a relatively short period of time of a few seconds. We show that the algorithm converges rapidly, with about half of the correction achieved by the first iteration. This has been observed in simulation and on-sky.
Furthermore, the correction of the NCPAs is effective even if it is considerably smaller than the dynamic of the atmosphere.

The characteristics of DrWHO are enumerated in section \ref{sec:Discussion}. The main strong points are the robustness of the algorithm, its independence from any model, making it compatible with other wavefront control methods, and that it does not rely on any linear assumption, thus particularly fit for a PyWFS.
In fact, some other techniques for compensating NCPAs mentioned in Section \ref{sec:intro} usually work independently from the WFS. They do their own correction based on a model, for example applying probes on a DM channel that is parallel to the AO loop DM channel. DrWHO combines focal plane wavefront sensing and the WFS. \\
Furthermore, there is a natural extension of this technique in high contrast imaging in the IR, to be more efficient by parallelizing the algorithm in terms of spatial frequencies, which would allow to correct speckles individually and to observe fainter targets.

This paper is the first proof of concept of the algorithm, and presents some first results of its PSF quality and stability enhancement. DrWHO shows a wide potential of improvement, which will be presented in future work.

\begin{acknowledgements}
Based [in part] on data collected at Subaru Telescope, which is operated by the National Astronomical Observatory of Japan. The development of SCExAO was supported by the Japan Society for the Promotion of Science (Grant-in-Aid for Research \#23340051, \#26220704, \#23103002, \#19H00703 \& \#19H00695), the Astrobiology Center of the National Institutes of Natural Sciences, Japan, the Mt Cuba Foundation and the director's contingency fund at Subaru Telescope. The authors wish to recognize and acknowledge the very significant cultural role and reverence that the summit of Maunakea has always had within the Hawaiian community. We are most fortunate to have the opportunity to conduct observations from this mountain. NS acknowledges support from the PSL Iris-OCAV project. NS and VD acknowledge support from NASA (Grant \#80NSSC19K0336). KA acknowledges support from the Heising-Simons foundation. We would like to thank the referee for their insightful comments. 
\end{acknowledgements}

\bibliographystyle{aa}
\bibliography{main}

\appendix
\section{Expanding on data synchronization and re-sampling}
\label{appendix}
\label{sec:syncdata}

\begin{figure*}[t]
\begin{center}
\centering
\includegraphics[width=0.8\textwidth]{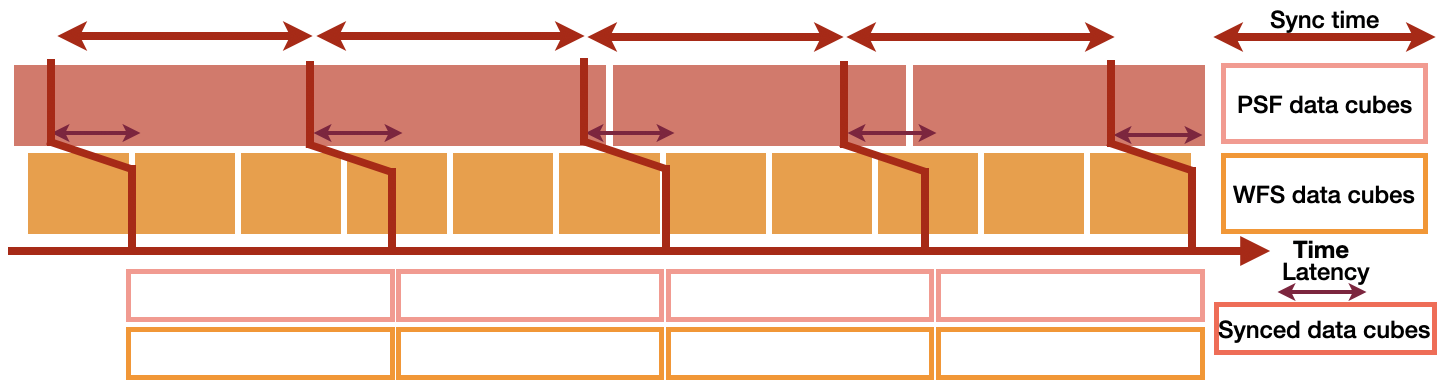}
\caption{Schematic view of how the synchronization of files works. Given two data streams with a known latency between them, the algorithm generates two synchronized files for those data.}
\label{fig:sync_files}
\end{center}
\end{figure*}

In is fundamental for DrWHO to synchronize and re-sample PyWFS and short exposure focal plane camera data streams. This is done directly on the telemetry files that are saved to the file system, but not in real time. Indeed, DrWHO needs frames to be synchronized but not precisely in real time, since one iteration of the algorithm is several seconds long. 
A script for telemetry synchronization is has been implemented, to which four inputs are required: 
\begin{itemize}
    \item the PyWFS data telemetry timeseries;
    \item the focal plane camera data stream;
    \item absolute timestamps of each exposure on both timeseries, retrieved from a hardware latency calibration process 
    \item the output sampling frame rate
\end{itemize}

For example, if the focal plane camera runs at 1kHz, the PyWFS camera runs at 2kHz, and a 3kHz output frame rate is requested, both input streams will be resampled to a common 3 kHz frame rate. Once launched, the synchronization script waits for data cubes (FITS files) and the corresponding timing files to appear on the hard drive, reads them and re-samples them to the same time baseline. This script handles differential latency between the streams, potential missed frames, and possible interruptions in the input streams acquisition. A linear interpolation is performed to re-sample input streams to output data cubes.
This script provides output data cubes of the required length (which is an input parameter of the script), that are saved on the file system. As described  above, the operation is not real-time as files need to appear to generate the re-sampling. Choosing the length of the data cubes is linked with the latency we need for DrWHO. The script needs to be started once only, then it catches up as the AO loop is running. Figure \ref{fig:sync_files} gives a visual explanation of this file synchronization.

\end{document}